\documentclass[iop]{emulateapj}

\usepackage{graphicx}
\usepackage{amsmath}
\usepackage{amssymb,latexsym,amsopn}

\shorttitle{FRB~131104}
\shortauthors{Ravi, Shannon \& Jameson}

\begin{document}

\title{A fast radio burst in the direction of the Carina dwarf spheroidal galaxy}

\author{V. Ravi\altaffilmark{1,2,3,4}, R. M. Shannon\altaffilmark{2}, and A. Jameson\altaffilmark{3}}
\affil{$^{1}$School of Physics, University of Melbourne, Parkville, VIC 3010, Australia}
\affil{$^{2}$CSIRO Astronomy and Space Science, Australia Telescope National Facility, PO Box 76, Epping, NSW 1710, Australia}
\affil{$^{3}$Swinburne University of Technology, Centre for Astrophysics and Supercomputing, Mail H39, PO Box 218, VIC 3122, Australia}

\altaffiltext{4}{E-mail address: v.vikram.ravi@gmail.com}

\begin{abstract}

We report the real-time discovery of a fast radio burst (FRB~131104) with the 
Parkes radio telescope in a targeted observation of the Carina dwarf spheroidal galaxy.
The dispersion measure of the burst is 779\,cm$^{-3}$\,pc, exceeding predictions for the maximum line-of-sight 
Galactic contribution by a factor of 11. The temporal structure of the burst is characterized by an exponential scattering tail with a timescale of 
2.0$^{+0.8}_{-0.5}$\,ms at 1582\,MHz that scales as frequency to the power $-$4.4$^{+1.6}_{-1.8}$ (all uncertainties represent 95\% confidence intervals). 
We bound the intrinsic pulse width to be $<0.64$\,ms due to dispersion smearing across a single spectrometer channel.  
Searches in $78$~hours of follow-up observations with the Parkes telescope reveal no additional sporadic emission  and no evidence for associated 
periodic radio emission. We hypothesize that the burst is associated with the Carina dwarf galaxy. Follow-up observations at other wavelengths 
are necessary to test this hypothesis. 

\end{abstract}


\section{Introduction}

The discovery of fast radio bursts (FRBs) presents a potentially transformational challenge to the understanding of the ephemeral Universe. 
FRBs are characterized by bright ($\sim$Jy peak flux densities) millisecond-duration pulses with 
inferred dispersion measures (DMs)  that significantly exceed (by factors of $3-20$) those expected from the Galaxy. 
Seven FRBs have recently been 
found in surveys at $\sim 1.4$\,GHz with both the Parkes \citep{lbm+07,tsb+13,bb14} and Arecibo 
\citep{sch+14} telescopes.

The intrinsic pulse durations, with some upper limits of  $\lesssim 1$~ms, imply  coherent emission originating from compact regions \citep{k14,lg14}.
 No additional  bursts were detected in  follow-up observations of one FRB \citep{lbm+07}, although these observations were conducted 
six years after the event. 
While FRBs have similar dispersion characteristics to the apparently  terrestrial  perytons \citep{bbe+11,kbb+12,sbm14,kon+14}, 
they can  be distinguished through their isolation on the sky, preferred locations away from the 
bulk of the Galactic disk \citep{bb14}, and broadband nature.

Any explanation for FRBs must account for the DM excesses, $\Delta {\rm DM}$, over the maximum line-of-sight DMs predicted by 
models for the Galactic electron density \citep{cl02}. 
Within the Milky Way, the large values of $\Delta {\rm DM}$ have been explained by invoking 
photo- or shock-ionized nebulae \citep{kon+14}, or emission from deep within stellar coronae \citep{lsm14}. 
However, these explanations are inconsistent, respectively, 
with constraints from H$\alpha$ observations on the presence of dense ionized nebulae \citep{kon+14}, and 
with the lack of observed deviations from the cold plasma dispersion law \citep{t14,k14b,d14}. 

If the Galactic electron density models are credible, and the frequency-dependent delays of FRBs are not intrinsic to their sources, 
it is possible that the DM excesses are caused by FRBs being extragalactic. 
In this case, the distances inferred from the values of $\Delta {\rm DM}$ are $\gtrsim1$\,Gpc in the absence of significant host contributions 
\citep{i03}.  A menagerie of exotic extragalactic sources of FRBs have been theorized, including  
  the collapse of gravitationally unstable magnetars to black holes \citep{fr14}, giant magnetar flares \citep{l14}, and emission from superconducting 
cosmic strings \citep{v08}. Despite some predictions \citep[e.g.,][]{rl14}, 
no astrophysical transient events at other wavelengths, such as supernovae or gamma-ray bursts, have yet been associated with FRBs.

The large positional uncertainties of $>3.5\arcmin$ at Arecibo and $>14\arcmin$ at Parkes have made it impossible to uniquely associate FRBs 
with any objects. 
Also, all FRBs have hitherto been detected in post-processing of data from blind 
radio surveys of large areas of the sky, and have therefore not been rapidly re-observed.

Here, we report the real-time discovery of a new FRB (131104) with the Parkes telescope.
The observations leading to the discovery of FRB~131104, which we detail in \S2, 
were targeted at the Carina dwarf spheroidal (Car~dSph) satellite of the Milky Way. An analysis of the temporal and 
spectral structure of the burst is presented in \S3. 
We  describe our immediate follow-up observations in \S4, discuss the implications of our results in \S5, and 
summarize in \S6. 


\section{Discovery observation}


We conducted our observations on UT 2013~November~4 (MJD~56600) with the $13$-beam 21\,cm multibeam (MB) receiver \citep{swb+96} at the prime focus of the 
64\,m Parkes antenna. 
The full-width, half-maxima of all beam responses on the sky are approximately 14\arcmin, although we note that 
the response patterns of the outer beams are mildly elliptical.
Data were recorded 
with the Berkeley-Parkes-Swinburne Recorder (BPSR) digital spectrometer using the same configuration used for the High Time Resolution Universe 
survey \citep{kjv+10}. 
 For each of the $13$ MB beams,  eight-bit 1024-channel spectra were integrated for 64\,$\mu$s intervals within a $400$\,MHz band centered on $1382$\,MHz. 
 Data from each of the linearly polarized feeds were summed to form total-intensity time series.

A real-time transient search pipeline, {\sc heimdall}\footnote{The source code for this pipeline is publicly available at 
http://sourceforge.net/projects/heimdall-astro.}  \citep{bbb+12}, was in operation on the BPSR computing 
cluster during the observations. 
This pipeline was used to search for isolated pulses at DMs between $1.5-2000$\,cm$^{-3}$\,pc 
with widths up to 0.262\,s. 
Events within our search ranges that were likely caused by radio-frequency interference (RFI)
were excised in real time \citep[see, e.g.,][]{bb10}. 
Specifically, events that were coincident in time with non-dispersed events or with ones in three or  more other beams 
were flagged as RFI. 
We tested the entire 
observing setup by pointing each beam in turn at the bright millisecond pulsar PSR J0437$-$4715 
to ensure that single pulses were being detected 
with the expected signal-to-noise ratios (SNRs).   We also carried out a 1\,hr observation of PSR B0540$-$69 in the 
Large Magellanic Cloud, which is the most distant known emitter of giant pulses \citep{jr03}. We detected one such pulse with a 
SNR of 7 at the known DM of 146.5. This detection is consistent with the measured rate of approximately two per hour.

\begin{figure}
\centering
\includegraphics[angle=-90,scale=0.38]{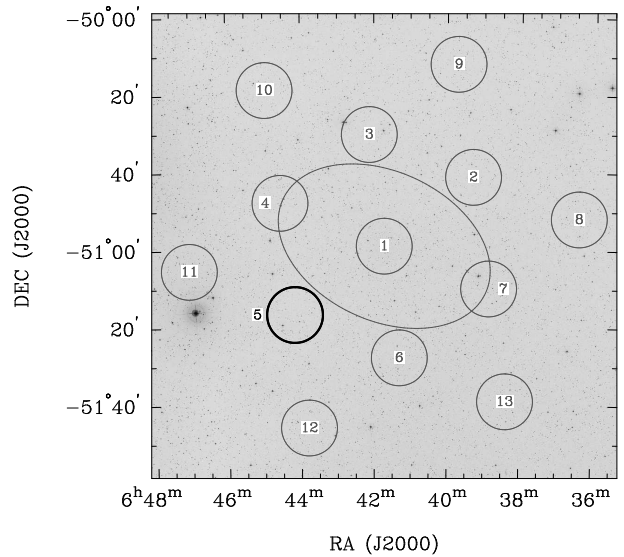}
\caption{The positions of the Parkes MB beams, labelled 1 to 13 and displayed as 14\arcmin~diameter circles, when FRB~131104 was detected. 
The FRB was only detected in beam 5 (thick black circle). 
The tidal ellipse of  Car~dSph is also shown \citep{ih95}, and the background inverted greyscale image is taken from the 
Second Digitized Sky Survey red images.}
\label{fig:map}
\end{figure}

\begin{figure}
\centering
\includegraphics[angle=-90,scale=0.36]{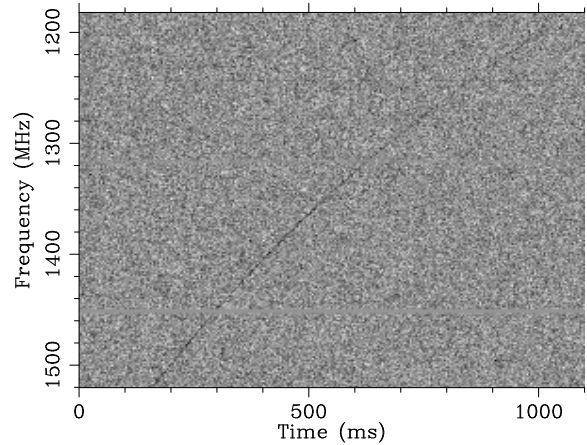}
\caption{Dynamic spectrum of FRB~131104, collapsed to a frequency resolution of 1.5625\,MHz and a time resolution of 4.1\,ms. 
The inverted greyscale intensity map represents the detected radio power in arbitrary units. 
Some persistent narrow-band RFI has been blanked.}
\label{fig:dyn}
\end{figure}

We then conducted a series of 1\,hr observations with the central MB beam positioned on Car~dSph. 
We disabled parallactic angle tracking for the MB feed system; consequently the outer MB beams rotated slowly on the sky. 
At 18:04:01.2 UT, approximately 21\,min following the start of the second observation of Car~dSph, the {\sc heimdall} pipeline reported the 
detection of a transient event at a DM of 779\,cm$^{-3}$\,pc, with a SNR of 30.6 when the time series was smoothed using a boxcar 
of width 2.08\,ms. The event was detected in beam 5 of the MB receiver, which at the time was pointed at the celestial coordinates (J2000) 
06h44m10.4s, $-$51d16m40s. The positions of all the beams on the sky when the event was detected are shown in Fig.~\ref{fig:map}. 
 Data recorded from beam 5 corresponding to the event   are shown as a dynamic spectrum in Fig.~\ref{fig:dyn}. 


We did not identify any significant events (with ${\rm SNR}>6$) in the other beams that were 
were coincident in time with the event in beam $5$. 
We also inspected 
the summed time series from all other beams dedispersed using the DM estimate from {\sc heimdall} of 779\,cm$^{-3}$\,pc, and again found 
no significant coincident occurrences. 
No coincident events were present when summed time series were formed using pairs of beams immediately 
surrounding beam 5.  

The background of significant candidate events with DMs greater than our threshold of 1.5\,cm$^{-3}$\,pc identified with the {\sc heimdall} pipeline 
during the discovery observation was characteristic of times not especially affected by RFI. These candidates typically had widths greater than 
100\,ms, and an approximately log-uniform distribution of DMs throughout the searched range. The event of interest was the highest-SNR occurrence during the 
discovery observation. 

The detected event is similar to the  FRBs of \citet{tsb+13}, \citet{sch+14}, and \citet{bb14}, and is markedly different from 
peryton emission \citep[][]{bbe+11}. 
Unlike perytons, and most forms of RFI, the event was isolated to a single beam of the MB receiver.
The DM estimate from {\sc heimdall} is much larger than the apparent DM of any published 
peryton, and exceeds the expected Galactic contribution along this line of sight by $\Delta {\rm DM}=710$\,cm$^{-3}$\,pc \citep{cl02}.  
An inspection of Fig.~\ref{fig:dyn} shows evidence for 
the presence of emission across the entire observed frequency band, in contrast to perytons, and the small derived width from   {\sc heimdall} of $\sim2$\,ms 
is also significantly lower than that of perytons. Further, while perytons predominantly manifest during the day \citep{bbe+11,kbb+12,sbm14}, our detection 
occurred at 05:04 local time, prior to sunrise. We hence conclude that the event is a FRB, designated  FRB~131104.

\section{Analysis of the temporal and spectral structure}

\begin{deluxetable*}{llllllllll}
\tabletypesize{\scriptsize}
\tablecaption{Details of time-frequency fits (all uncertainties are the 95\% confidence intervals).}
\tablewidth{0pt}
\tablehead{
\colhead{Model} & \colhead{$c_{1232}$ (Jy\,ms)} & \colhead{$c_{1332}$ (Jy\,ms)} & \colhead{$c_{1432}$ (Jy\,ms)} & \colhead{$c_{1532}$ (Jy\,ms)} & \colhead{DM (cm$^{-3}$\,pc)} & \colhead{$\tau_{s}$ (ms)} & \colhead{$\alpha$} & \colhead{$t_{0}$ (s)} &  \colhead{$\Delta$BIC}
}
\startdata
0 & 1.9$^{+0.2}_{-0.2}$ & 2.0$^{+0.1}_{-0.2}$ & 2.7$^{+0.2}_{-0.2}$ & 3.7$^{+0.4}_{-0.5}$ & 779.1$^{+0.1}_{-0.2}$ &  $-$ & $-$ & 1.03435$^{+7\times10^{-5}}_{-8\times10^{-5}}$ & 0 \\
1 & 2.3$^{+0.2}_{-0.3}$ & 2.0$^{+0.2}_{-0.1}$ & 2.3$^{+0.1}_{-0.2}$ & 3.0$^{+0.3}_{-0.4}$ & 779.0$^{+0.2}_{-0.2}$ &  1.1$^{+0.1}_{-0.1}$ & $-$ & 1.0338$^{+1\times10^{-4}}_{-1\times10^{-4}}$ & -299 \\
2 & 2.5$^{+0.4}_{-0.3}$ & 2.0$^{+0.2}_{-0.2}$ & 2.2$^{+0.1}_{-0.2}$ & 2.6$^{+0.4}_{-0.4}$ & 778.5$^{+0.2}_{-0.3}$ & 2.0$^{+0.8}_{-0.5}$ & 4.4$^{+1.6}_{-1.8}$ & 1.0340$^{+1\times10^{-4}}_{-1\times10^{-4}}$ & -306 
\enddata
\label{table:t1}
\end{deluxetable*}

\begin{figure}
\centering
\includegraphics[angle=-90,scale=0.62]{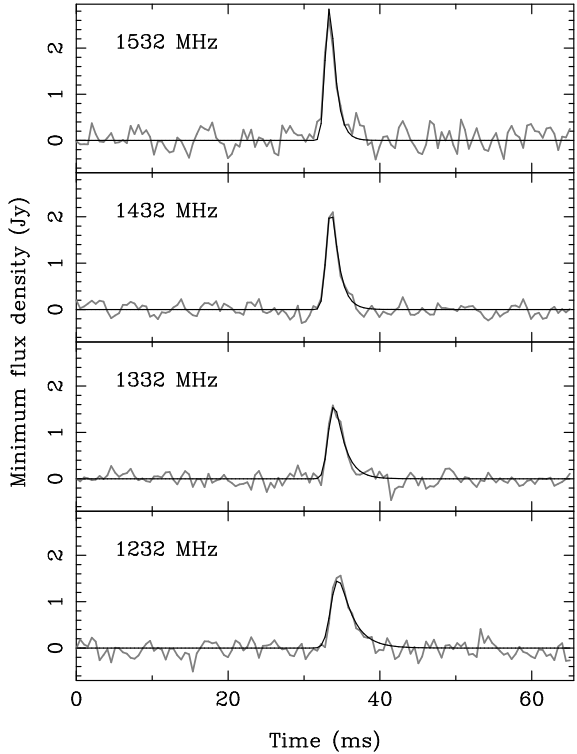}
\caption{Comparison between the best-fitting model (black) and the frequency-dependent profile (gray) of FRB~131104. The data were dedispersed at the 
best-fit DM listed in Table~\ref{table:t1}, split into 100\,MHz bands centered on the displayed frequencies, 
and summed to a time-resolution of 0.512\,ms. 
The flux density scale is based on our on-sky noise temperature measurements for beam 5 of the Parkes MB receiver.
}
\label{fig:model}
\end{figure}

The temporal profile of FRB~131104, displayed in Fig. \ref{fig:model}, shows evidence for changes in both amplitude and width as functions of the radio frequency.  
In contrast to the modeling of the frequency-dependent delays and shapes of previous FRBs \citep{lbm+07,tsb+13,sch+14,bb14}, we conducted  a 
Bayesian analysis of the total intensity time series to measure the burst physical parameters. This approach allowed us to fully account for covariances between model parameters, and to obtain accurate confidence intervals for our parameter estimates. We used 
the Bayes Information Criterion \citep[BIC;][]{s78} to perform model selection and justify the successive inclusion of free parameters. 
The BIC for a given model is $-2\ln{\hat{L}}+k[\ln(n)-\ln(2\pi)]$, where $\hat{L}$ is the likelihood estimate, $k$ is the number of model parameters, 
and $n$ is the number of measurements. If adding a new free parameter to a model reduces the BIC, the new model is accepted because 
the likelihood increase is not simply due to the addition of the new parameter.  

In order to accelerate the analysis, the data were divided into 16 sub-bands across the 
full 400\,MHz band.  The two uppermost in frequency were discarded because instrumental filtering of 
RFI from the \textit{Thuraya} 3 satellite \citep{kjv+10} rendered them unusable. The FRB was detectable by eye in all other bands. We then dedispersed the data in each of these remaining bands 
using the initial DM estimate of 779\,cm$^{-3}$\,pc to form $14$ time series, while retaining the dispersion delays between each band. Adjacent samples in 
each time series were summed to a time-resolution of $0.512$\,ms. Finally, through visual inspection of these time series, we selected windows 
of width $15.36$\,ms containing all the evident signal, and used these measurements as the basis for our analysis. 

We assessed the noise properties of the time-series by examining $2$\,s of data immediately adjacent to these windows in each frequency sub-band.
We  found that the  values  
were consistent with  normal distributions.    We therefore measured the variances of these time series and assumed that the noise during the $15.36$\,ms 
windows was statistically identical.

We considered a hierarchical series of models for the pulse morphology.  
The simplest model consisted of a Dirac delta function impulse, dispersion-smeared at each frequency over a bandwidth corresponding to 
the BPSR frequency resolution of 0.391\,MHz. We modeled the smeared pulse profile as a Gaussian function, which approximately corresponds to the frequency 
response of the two-tap digital filterbank, and made the conventional assumption of cold plasma 
dispersion.
In this case, the signal intensity  $S$ at a given time $t$ and frequency $\nu$ was therefore:
\begin{equation}
\label{eq:1}
S(t,\nu) = \frac{A(\nu)}{\sqrt{2\pi\sigma_{\rm DM}^{2}(\nu)}}{\rm exp}\left[\frac{-(t-t_{0}-\tau_{\rm DM}(\nu))^{2}}{\sigma^{2}_{\rm DM}(\nu)}\right],
\end{equation}
where $t_{0}$ is a reference time at a reference frequency of $\nu_{0}=1582$\,MHz, 
\begin{equation}
\label{eq:2}
\tau_{\rm DM}(\nu)=(4.15\,{\rm ms}){\rm DM}[(\nu/{\rm GHz})^{-\beta}-(\nu_{0}/{\rm GHz})^{-\beta}]
\end{equation}
with $\beta=2$, and
\begin{equation}
\label{eq:3}
\sigma_{\rm DM}(\nu)=(1.622\times10^{-3}\,{\rm ms}){\rm DM}(\nu/{\rm GHz})^{-\beta-1}.
\end{equation}
Finally, we modeled the fluence, $A(\nu)$, in four separate 100\,MHz bands within the full observing bandwidth; in order of increasing frequency, we denote 
these fluences by $c_{1232}$, $c_{1332}$, $c_{1432}$, and $c_{1532}$.  
We evaluated the six free parameters of this model by exploring the likelihood space using the \textit{emcee} Markov Chain Monte Carlo (MCMC) 
software package \citep{fhl+13}. In the first row of Table~\ref{table:t1} we present the maximum-likelihood parameter values  and their 95\% confidence intervals for this model, which we term Model~0.
 
While in our fitting procedure we expressed the fluences as SNRs in 
each sub-band, we present them in physical units of Jy\,ms in Table~\ref{table:t1}. The conversions are based on our on-sky noise temperature 
measurements for beam 5 of the Parkes MB receiver using the unresolved radio source Hydra~A, and under the assumption that 
the FRB originated at the boresight of the beam. Our fluence measurements are therefore  lower limits.

We extended this model by convolving the Gaussian function in Equation~\ref{eq:1} with a one-sided exponential with timescale 
\begin{equation}
\label{eq:4}
\tau(\nu)=\tau_{s}(\nu/\nu_{0})^{-\alpha}, 
\end{equation}
where we initially assumed $\alpha=0$, and where $\tau_{s}$ is a free parameter.   This allows for the pulse to be modeled with an exponential tail. 
This model, denoted by Model~1 in Table~\ref{table:t1}, significantly reduced the BIC with respect to Model~0. A model where we instead 
replaced  $\sigma_{\rm DM}(\nu)$ in Equation~\ref{eq:1} with $[\sigma^{2}+(\sigma_{\rm DM}(\nu)-\sigma_{\rm DM}(\nu_{0}))^{2}]^{1/2}$, where 
$\sigma$ is a free parameter, did not produce as great a reduction in the BIC. A model which instead included $\beta$ as a free parameter also 
did not produce as great a reduction in the BIC. 

We then allowed $\alpha$ to be a free parameter in addition to $\tau_{s}$, enabling a search for scattering as an origin 
for the exponential tail. This caused a further reduction in the BIC. The resulting parameters are also listed in Table~\ref{table:t1}, 
and denoted by Model~2. Neither the addition of $\sigma$ nor the addition of $\beta$ as a free parameter further reduced the BIC. We hence have 
no evidence for a pulse half-width, $\sigma$, at $\nu=\nu_{0}$ that is different to the dispersion smearing timescale $\sigma_{\rm DM}(\nu_{0})=0.32$\,ms. 
We also have no evidence for $\beta \neq 2$. 

In constrast, we find strong evidence for frequency-broadening of the pulse profile. The broadening timescale is 
$\tau_{s}=2.0^{+0.8}_{-0.5}$\,ms at $1582$\,MHz, and the index of the frequency-dependence is $\alpha=4.4^{+1.6}_{-1.8}$ (all uncertainties are the 95\% confidence intervals). 
This is the second time that a FRB has been shown to exhibit such an exponential tail \citep[after FRB~110220;][]{tsb+13}, although 
other studies have found that FRB effective widths have similar frequency dependencies \citep{lbm+07,tsb+13,bb14}.  
The value of $\alpha$  corresponds to scattering by a Kolmogorov-turbulent plasma \citep[in which case $\alpha=4.4$ 
is expected; e.g.,][]{r77}, and is consistent with both  scattering in a diffuse, extended medium, or in a geometrically thin region along the line of sight, or both. 


Based on the measurements of the fluences,  we find $A(\nu)\propto \nu^{0.3\pm0.9}$.
This marginally inverted spectrum is broadly consistent with the analysis  
of FRB~121102 detected with the Arecibo telescope by \citet{sch+14}. 
We note that, depending on the location of the FRB within the primary beam, both hardening and 
softening of the spectrum can be induced.

\section{Follow-up observations}

We re-observed the field of FRB~131104 for three hours immediately following its occurrence, continuously rotating the receiver to keep beam~$5$ 
directed toward the FRB position. 
We also observed the field for 6\,hr lengths during the subsequent three days, and for an additional 57\,hr at arbitrary times during the 
following year.  With the exception of RFI, we did not detect any other event 
with ${\rm DM}>1.5$\,cm$^{-3}$\,pc, a width of $\leq65$\,ms, and ${\rm SNR}>8$ in any of the MB beams. The upper width threshold aided in rejecting the 
majority of spurious candidates. 

Given the all-sky FRB rate estimated by \citet{tsb+13},  we would have expected to see 0.5 FRBs in the totality of our observations. 
Assuming that the occurrence of FRBs can be modeled as an isotropic Poisson process, the probability of finding one or more 
FRBs in all our observations is 0.4.

We also searched for periodic radio emission in the FRB discovery observation, and in four 0.5\,hr follow-up observations of the apparent FRB position 
spaced evenly over six months. We used standard pulsar search software to search both at the DM of the FRB and at all DMs $>1.5$.  
No periodic signal was found with ${\rm SNR}>8$ and a period between 1\,ms and 10\,s that was not attributable to RFI. For a long-period pulsar with a duty cycle of 
10\%, the limiting sensitivity of our search at each epoch was 0.1\,mJy.

\section{Discussion}

FRB~131104 is the first FRB to be found in a targeted search for such events. 
Car~dSph, with a moderate Galactic latitude ($b=22.2^{\circ}$), is unique among dwarf Milky Way companions in having undergone three widely 
spaced episodes of star formation \citep{hmn98}, and the oldest stellar population shows some evidence for tidal 
disruption \citep{bit+12,mbl+14}.  We chose to observe  Car~dSph because of the low expected 
Galactic contribution to the DM in its direction,
the presence of multiple stellar populations and possible tidal debris, the fact that the \citet{lbm+07} FRB was detected near the Small Magellanic Cloud, and the possibility of discovering pulsars 
in Car~dSph in order to identify the Local Group contribution to DM along this sightline.  
Despite the inferred FRB rate of $10^{4}$\,sky$^{-1}$\,day$^{-1}$ \citep{tsb+13} and the total effective Parkes field-of-view of 
$\sim2\times10^{-5}\times4\pi$\,str, this event remarkably occurred within 1.5\,hr of the beginning of our observations of Car~dSph. 

The location of beam 5 of the MB receiver when FRB~131104 was detected (Fig.~\ref{fig:map}) is also coincident with stellar debris associated with 
the Large Magellanic Cloud \citep{mbl+14}, which is a factor of $\sim2$ closer to the Earth than Car~dSph. However, the proximity of the half-power 
beam point of beam 5 to the Car~dSph tidal ellipse means that FRB~131104 may indeed be coincident in sky location with that galaxy. 

However, for FRB~131104 to have originated within  Car~dSph,  which is  $101\pm5$\,kpc  distant from the Solar system \citep{m98},  its source 
would have to be located behind, or embedded in, an overdensity of ionized interstellar gas relative to the Milky Way ionized halo. 
Recent estimates for the electron density of the Milky Way ionized halo 
suggest $n_{e}=(2\pm0.6)\times10^{-4}$\,cm$^{-3}$ \citep{gmk+12}.  Even if the  
DM contribution from Car~dSph  is 100\,cm$^{-3}$\,pc, the intervening medium would need to have a mean electron density of 
$n_{e}\sim6\times10^{-3}$\,cm$^{-3}$. 

Such large densities have however been measured in the Magellanic stream \citep[e.g.,][]{fwb+14}, a part of which lies 
along the Car~dSph sightline. FRB~010724 \citep{lbm+07}, observed just a few degrees South of the Small Magellanic Cloud, 
had a sightline that was clearly associated with the Magellanic stream. 
Further detailed observational analyses of the ionized gas content surrounding the Milky Way 
along the line of sight to FRB~131104 would reveal whether there is sufficient material to associate this FRB with Car~dSph.

Pending such investigations, we cannot constrain the distance to FRB~131104. Hence, various possibilities for the origin of FRB~131104 remain open. 
If FRB~131104 originated from the close surrounds of the Milky Way, it could conceivably  represent a form of giant pulse emission. 
The brightest such event detected from the Crab pulsar with the Arecibo telescope at 0.43\,GHz 
would have had a SNR of $10^{6.5}$ in the absence of radio emission from the Crab nebula \citep{cbh+04}. Assuming a flux density spectrum 
$\propto \nu^{-3}$ for giant 
pulses, such an event would have been detectable with Parkes at the distance to Car~dSph with a SNR of $\sim26$. Furthermore, the energy distribution 
of giant pulses from the Crab pulsar appears to flatten at the high fluence end \citep{cbh+04,mml+12}, suggesting that even brighter 
pulses may yet be found than the \citet{cbh+04} event. 
Such a scenario for FRB~131104 will be best confirmed through an independent estimate of the line-of-sight DM to  Car~dSph, and 
through the observation of a repeat event. 

\section{Summary}

FRB~131104 was detected in a targeted observation of the Milky Way satellite Car dSph with the Parkes radio telescope.
It is possible that the large dispersion of the burst
may be partly associated with circum-Galactic ionized gas along the line of sight.  
The pulse was found to have both an an exponential tail and an inverted spectrum;  the frequency dependence of the tail is consistent with scattering in a turbulent medium.
The real-time detection of FRB~131104 enabled rapid follow-up with the Parkes telescope, which 
revealed no sporadic or periodic radio emission.  Further analyses of circum-Galactic ionized gas along the sightline to 
FRB~131104 are required to constrain the distance and origin of this FRB.

\acknowledgments

We thank M.~Bailes and E.~Petroff for useful discussions, and the Swinburne pulsar group for making their real-time single pulse detector
available for this experiment. We also thank the  Parkes Observatory staff for hospitality during our visits.
The Parkes Radio Telescope is part of the Australia Telescope, which is funded by the Commonwealth of Australia for operation as a National Facility by CSIRO.
We acknowledge the use of NASA's {\em SkyView} facility
     (http://skyview.gsfc.nasa.gov) located at NASA Goddard Space Flight Center. 
        This work was performed on the {\textit gSTAR}  national facility at Swinburne University of Technology.   {\textit gStar}   is funded by Swinburne and the Australian Government's Education Investment Fund.

\end{document}